\documentclass[preprint2]{aastex}

\slugcomment{}

\shorttitle{X-ray observations of a z=6.30 QSO}
\shortauthors{Farrah et al.}

\begin{document}

\title{The X-ray Spectrum of the $z=6.30$ QSO SDSS J1030+0524}


\author{D. Farrah}
\affil{Spitzer Science Center, Jet Propulsion Laboratory, California Institute of Technology, Pasadena, CA 91125, USA}
\author{R. Priddey}
\affil{Department of Physics, Astronomy and Mathematics, University of Hertfordshire, UK}
\author{R. Wilman}
\affil{University of Durham, UK}
\author{M. Haehnelt \& R. McMahon}
\affil{Institute of Astronomy, University of Cambridge, Cambridge, UK}



\begin{abstract}
We present a deep XMM-Newton observation of the $z=6.30$ QSO SDSS J1030+0524, the second most distant quasar currently known. The data contain sufficient counts for
spectral analysis, demonstrating the ability of XMM-Newton to measure X-ray spectral shapes of $z\sim6$ QSOs with integration times 
$\gtrsim100$ks. The X-ray spectrum is well fit by a power law with index $\Gamma=2.12\pm0.11$, an optical-X-ray spectral 
slope of $\alpha_{ox}=-1.80$, and no absorption excess to the Galactic value, though our data are also consistent with a 
power law index in the range $2.02 < \Gamma < 2.5$ and excess absorption in the range $0 < {\rm N_{H}(cm^{-2})} < 8\times10^{22}$. 
There is also a possible detection ($\sim2\sigma$) of FeK$\alpha$ emission.  The X-ray properties of this QSO are, overall, 
similar to those of lower-redshift radio-quiet QSOs. This is consistent with the statement that the X-ray properties of 
radio-quiet QSOs show no evolution over $0<z<6.3$. Combined with previous results, this QSO appears indistinguishable in 
any way from lower redshift QSOs, indicating that QSOs comparable to those seen locally existed less than one Gyr 
after the Big Bang.  
\end{abstract}



\section{Introduction} \label{intro}
One of the most important goals in modern cosmology is to understand the formation of galaxies and large-scale structures. 
According to the currently prevalent theory for structure formation, biased hierarchical buildup within the $\Lambda$CDM framework (hereafter 
referred to as $\Lambda$CDM), the growth of large galaxies results from mergers between smaller systems, with more massive galaxies hosted 
within more massive dark matter halos, themselves the rarest, most biased overdensities in the underlying mass distribution. Observations of 
QSOs and other massive systems are thus an excellent way of testing structure formation models. This is particularly apposite for QSOs 
at very high redshifts, where the added constraint of the youth of the cosmos allows for the most stringent tests. The discovery \citep{fan01,fan03} 
from the Sloan Digital Sky Survey of QSOs at $z>6$, when the Universe was (under any currently favored cosmology) less than one Gyr old, presents the 
opportunity to perform such tests. The first step is to determine the nature of the central engines in these QSOs, as $\Lambda$CDM models 
generally extrapolate the host halo mass from the central black hole mass \citep{hnr98}. This is readily achieved using X-ray observations, 
which directly probe the central engines of QSOs, largely free of obscuration bias. Deep X-ray observations of $z>6$ QSOs, to measure their X-ray 
spectral shapes and luminosities, can therefore be used to examine the properties of their central black holes, and to compare to the X-ray 
properties of lower redshift QSOs to search for drivers behind the strong evolution of the QSO luminosity function \citep{boyl00}. 

One such QSO is SDSS J1030+0524 at 10$^{h}$30$^{m}$27$^{s}$.1 +05$^{d}$24$^{m}$55$^{s}$.0 (J2000), with a redshift of $z=6.30$ \citep{ric04}. 
Deep imaging observations show that this QSO is not significantly lensed \citep{ric04} or beamed \citep{hce02}, and that the rest-frame 
optical continuum shape and luminosity are probably similar to those of lower redshift QSOs \citep{fan01}. The rest-frame UV spectrum shows an almost complete 
Gunn-Peterson trough, indicating that this QSO lies within the epoch of reionization \citep{bec01}. Further observations found 
evidence for supersolar metallicities \citep{pen02,fre03} and indirect evidence for a formed host galaxy \citep{blo03}. Although not detected 
in the sub-mm \citep{pri03}, the radio \citep{pet03} or the ROSAT all-sky survey, this QSO is detected in Chandra snapshot observations \citep{bra01}. 
In this letter, we present deep XMM-Newton observations of SDSS J1030+0524.  We assume $\Omega=1$, $\Lambda=0.7$ and H$_{0}=70$ km s$^{-1}$ Mpc$^{-1}$.


\section{Observations \& analysis} \label{obs}
SDSS J1030+0524 was observed continuously by XMM-Newton for 105ks in May 2003. The source was placed on axis, and observed by all 
the onboard instruments, although we only consider data from the European Photon Imaging Counter (EPIC) Metal Oxide Semiconductor (MOS) and pn 
cameras here. The data were reduced using the XMM Science Analysis System (SAS) v6.0 package. The event lists were filtered to include 
only single and double events (patterns 0-4 for the pn and 0-12 for the MOS) with quality flag 0, and to remove time periods during which 
the background was excessively high due to proton flares. Light curves were extracted using the whole MOS and pn fields of view to check that 
all the flaring background periods had been removed. The resulting effective exposure time was $\sim75$ks. The source spectrum was extracted 
using a circular aperture of radius 45\arcsec\ centered 
on the source, and the background spectrum was extracted from a contiguous region free from sources and with the same 
instrumental background as our target. Additional light curves were extracted from the same region as the source spectra to search for 
short-term variability from the QSO. Spectral analysis was performed using the Xspec v11.3 package \citep{arn96}, using spectra binned to 
have at least 20 counts per bin.

\section{Results} \label{res}
Before undertaking detailed analysis, we checked for consistency between the three EPIC detectors by fitting the MOS and pn spectra 
with single power-law models. The derived photon indices and normalizations agreed to within $0.25\sigma$. We therefore 
proceeded to fit the three spectra simultaneously. 

\begin{figure}
\includegraphics[angle=0,scale=0.32]{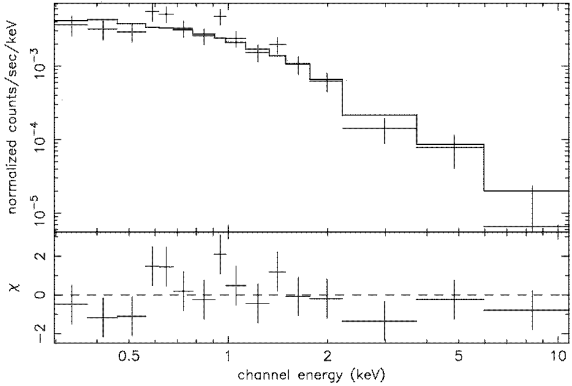}
\caption
{
EPIC pn spectrum of SDSS J1030+0524, plus a single power law fit \& fit residuals with absorption fixed at the Galactic value. The MOS 
spectra have not been plotted for clarity. \label{spectrum}
}
\end{figure}

The X-ray spectrum of SDSS J1030+0524 is presented in Figure \ref{spectrum}. The final spectrum contained $\sim560$ counts in total, where 
340 were detected by the pn camera and 220 by the two MOS cameras. A summary of the results from spectral fitting is given in Table 
\ref{specfit}. For all fits we assumed a Galactic hydrogen column of 3.2$\times10^{20}$ cm$^{-2}$ \citep{sta92}. We first considered a single 
power law model. This gave a reasonable fit to the combined spectrum with a photon index of $\Gamma=2.12\pm0.11$. We speculate that the excess flux 
between 0.9-1.0 keV may arise from a complex of iron emission features (6.4-7.0 keV rest-frame), however the flux in this bin 
is only 2.1$\sigma$ above the power law fit; we cannot therefore confirm or refute this without higher quality data. We also derived the value 
of $\alpha_{ox}$, the slope of a nominal power-law between 2500\AA\ and 2KeV. Using both the 2500\AA\ flux and method to compute $\alpha_{ox}$ 
described by \citet{bra01}, together with our X-ray data, yields $\alpha_{ox}=-1.80$. No short term variability in the X-ray emission was 
detected in any band, though given the low count rates in both EPIC detectors this is not unexpected.  

\begin{figure}
\includegraphics[angle=0,scale=0.32]{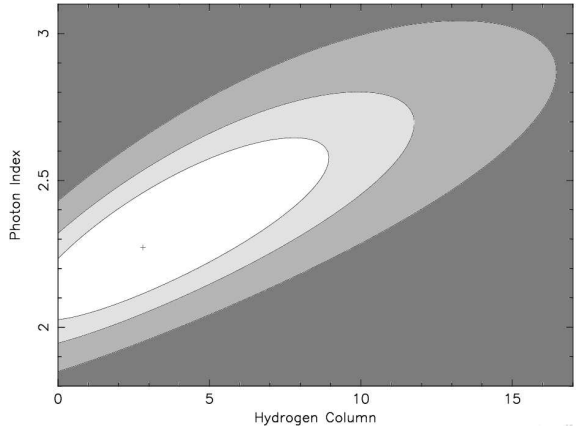}
\caption
{
Confidence plot of $\Gamma$ vs. $N_{\rm H}$ (units are $10^{22}$ cm$^{-2}$) from the fit where both parameters are allowed to vary. 
Contours are 1$\sigma$, 2$\sigma$ \& 3$\sigma$. \label{conf}
}
\end{figure}

Previous studies (\citet{rtu00}, though see also \citet{vig03}) have found possible evidence for increasing absorption with increasing 
redshift in QSO X-ray spectra, rising from $\sim10^{20}$cm$^{-2}$ at $z\sim0.1$ to $\gtrsim10^{22}$cm$^{-2}$ at $z\gtrsim2$. It is plausible 
therefore that the X-ray 
emission from SDSS J1030+0524 is attenuated by absorption local to the QSO. A modest amount of excess absorption might be 
expected on the grounds that the Ly$\alpha$ emission shows some HI absorption \citep{blo03}, although if the absorption is due to shocked material 
accreting onto the host galaxy halo it may not have sufficient heavy element content to produce appreciable X-ray absorption. We consider three 
power law plus absorption scenarios; a model where both $\Gamma$ and $N_{H}$ vary, a model where the power law slope is fixed 
at $\Gamma=2$, and a model where the hydrogen column is fixed at $N_{H}=10^{22}$cm$^{-2}$. In the case where $N_{H}$ is fixed 
the derived photon index, at $\Gamma=2.27\pm0.2$, lies within $1\sigma$ of the photon index derived assuming zero intrinsic absorption. In the two 
cases where $N_{H}$ is allowed to vary the best-fit values of $N_{H}$, although statistically consistent with zero, have errors too large 
to allow an accurate hydrogen column to be quoted. We can however explore the range of 
acceptable values. A confidence plot of $\Gamma$ vs. hydrogen column for the fit in which both $\Gamma$ and $N_{H}$ are allowed to vary is shown in 
Figure \ref{conf}. Based solely on this plot, we would quote 1$\sigma$ ranges of $2.02 < \Gamma < 2.6$ and $0 < N_{H}$(cm$^{-2}) < 9\times10^{22}$. 
A hydrogen column above  $\sim8\times10^{22}$cm$^{-2}$ however requires $\Gamma>2.5$; this is significantly higher than for any other observed QSO 
(see e.g. \citet{rtu00}), though such steep slopes are seen in Seyferts \citep{wfi93}. We therefore quote the `best fit' parameters as $\Gamma=2.12\pm0.11$, 
with no evidence for excess absorption, but also quote approximate acceptable ranges on these two parameters that are consistent 
with our data, these are $2.02 < \Gamma < 2.5$ and $0 < N_{H}$(cm$^{-2}) < 8\times10^{22}$.

We also explored a wider range of models for the X-ray emission from this QSO, though given the relatively low quality of our data we advocate 
caution in interpreting the following results, and do not present them in Table \ref{specfit}. A single thermal bremsstrahlung or blackbody model 
are both substantially worse fits than any of the power law models, and can be rejected. A power law plus thermal bremsstrahlung model on the 
other hand, with or without local absorption, gives a fit of comparable quality to the power law model but requires an X-ray spectral slope of 
$\Gamma\sim1.45$; this is much flatter than the spectral slopes of other radio-quiet QSOs \citep{rtu00}. While we cannot formally discount the 
possibility of significant thermal X-ray emission, we consider it unlikely. A power law $\times$ reflection model gives a marginally worse fit 
than a single power law model, but not sufficiently so to allow us to formally reject a model including reflection. A power law model with a 
gaussian line at an observed-frame energy of $\sim0.95$KeV gives a somewhat better fit than a single power law model (with $\chi^{2}_{red}/$DOF$=1.4/23$) 
but this improvement in fit quality is not sufficient to formally prefer this model over the single power law model. Furthermore, the gaussian line 
properties are not well constrained; the only constraint is that the line width must be less than 600eV at 3$\sigma$ confidence. Overall, 
we conclude that the X-ray spectrum of SDSS J1030+0524 is best-fit by the single power law model, but we cannot determine the presence or otherwise 
of an FeK$\alpha$ line, and cannot rule out contributions from a thermal bremsstrahlung or reflected component. 

Finally, we note that the single power law model, whilst statistically acceptable, is not a `good' fit to the data, with a null hypothesis 
probability of $6\%$. Inspection of the residuals in Figure \ref{spectrum} shows that this is primarily due to several `features' at observed-frame 
energies $<1$KeV, which are only apparent in the pn spectrum as the MOS spectra contain significantly fewer counts. These features (which 
include the possible FeK$\alpha$ line) have three possible origins; they could be intrinsic to the QSO, they could arise from foreground (i.e. 
$z<6.3$) sources, or they could be residuals from (for example) soft proton flares that were not fully removed by the data reduction. Of these 
three reasons the first two are the most likely, but we cannot examine the reasons behind the relative poorness of fit further without higher 
quality data. 

\section{Discussion} \label{disc}
The advent of XMM-Newton and Chandra have revolutionized X-ray studies of high redshift QSOs. Currently, detailed X-ray spectral information 
is available for a large sample of QSOs in the redshift range $0<z<5$ \citep{lao97,rtu00,ree01,fbr03,vig03,gru04}. We first compare results 
from these studies to the X-ray properties of SDSS J1030+0524. Our best-fit spectral index, at $\Gamma=2.12\pm0.11$, lies near the middle of 
the range of values of $\Gamma$ found for lower redshift radio quiet QSOs (RQQs) studied with ASCA \citep{rtu00}, and is 
comparable to the values of $\Gamma$ derived for $z\sim4$ RQQs from XMM-Newton observations \citep{fbr03,gru04}. The derived rest-frame 
luminosity (which has a total $1\sigma$ uncertainty of  $\sim15\%$) is approximately 1.5 times and 3 times higher than those derived from 
Chandra observations by \citet{mat02} and 
\citet{bra01} respectively (who assume $\Gamma=2.0$). We attribute this difference to the lower S/N of the Chandra data rather than the difference 
in values of $\Gamma$, though variability is also a possibility, and one that our data do not allow us to test for. Our 2-10keV luminosity 
is however comparable to lower redshift RQQs \citep{rtu00} 
and marks SDSS J1030+0524 as being no more than averagely luminous in the X-ray.  Our value for the optical-X-ray spectral slope, $\alpha_{ox}=-1.80$, 
is well within the observed range for lower-redshift RQQs, and is statistically identical to the mean value of $\alpha_{ox}$ for $z\sim4$ RQQs 
\citep{vig03a}. Overall therefore, the X-ray luminosity and spectral shape of SDSS J1030+0524 appear to be indistinguishable from those of 
RQQs at $0<z<5$ (see e.g. Fig. 8 of \citet{vig03}). Although statistics based on one object are obviously not trustworthy, this result is 
consistent with the statement that the X-ray properties of optically selected RQQs show no evolution up to $z=6.3$. The only exceptions would be 
if (1) the unlikely event that this QSO has a very steep intrinsic X-ray spectrum attenuated by heavy absorption, such systems are
rare amongst the radio-quiet QSO population, and (2) if the marginal excess flux in the 0.9 - 1.0keV energy range really is FeK$\alpha$ emission, 
as this is very rare in QSOs generally. Iron K emission from QSOs and Seyferts is thought to arise either from a cold reflected 
component from the accretion disk (for line widths $\sim50$eV), or from hot gas in a halo (for line widths $\gtrsim100$eV, in this case we might 
also expect to see significant thermal X-ray emission). The detection 
of Iron K lines in SDSS J1030+0524 would therefore be especially interesting, but given the quality of our data we do not consider this further. 

We can combine this result with previous studies of SDSS J1030+0524 to establish whether its global properties differ from lower redshift 
RQQs. As described previously, this QSO does not appear to be significantly magnified by lensing \citep{ric04}. Together with its measured 
X-ray luminosity and the spectral shape, this constitutes compelling evidence that the mass estimate of $\sim2\times10^{9}$M$_{\odot}$ for 
the central black hole derived from the 1450\AA\ magnitude \citep{fan01} is accurate. Considered together with previous observations (see \S\ref{intro} 
for a review) then SDSS J1030+0524 appears to be indistinguishable in {\it any} way from the lower redshift RQQ population. The very existence 
of such a system only 860Myr (under our cosmology) after the Big Bang poses a significant challenge for structure formation theories, and it 
is this theme we explore in the remainder of this discussion. 

To explain the existence of SDSS J1030+0524 requires the formation of a $\sim10^{9}$M$_{\odot}$ black hole and probably also a $\sim10^{14}$M$_{\odot}$ 
DM halo (e.g. \citet{hnr98}) and $\sim10^{11}$M$_{\odot}$ of stars \citep{mag98}, all within 860Myr. The formation of a suitably massive halo is 
readily achieved within the $\Lambda$CDM framework \citep{mwh02}, and simulations predict that the mass profiles in the inner 10$h^{-1}$kpc of 
the most massive halos evolve very little at $z\lesssim6$ \citep{fma01,gao04}. These simulations however only consider the hydrodynamical evolution 
of the dark matter distribution, and do not consider the astrophysical processes of star and black hole formation. The formation of the stars and 
central black hole must therefore be considered separately. 
There is evidence, both from observations of old ellipticals at $z=1.5$ \citep{pea98} and from the observed upper bound on the 
line-of-sight velocity dispersions of stars in ellipticals \citep{lpe03}, that $\Lambda$CDM must be capable of forming the stars 
in a giant elliptical galaxy in less than one Gyr. This requires (for example) a `burst' of 1000M$_{\odot}$ 
yr$^{-1}$ star formation lasting $\sim$100Myr. Such high instantaneous star formation rates are inferred to exist in the $1<z<4$ 
sub-mm survey sources (e.g. \citet{bor03}), and in $z\gtrsim4$ QSOs \citep{isa02}, and 100Myr is a reasonable upper age limit 
for starbursts based on observations of local starbursts \citep{far03}. To form the host galaxy of SDSS J1030+0524 therefore 
appears feasible, though this inference is based on observations of lower redshift systems, and we note that the upper limit on the 
sub-mm flux from SDSS J1030+0524 \citep{pri03} implies an upper limit on the instantaneous star formation rate of $300$M$_{\odot}$yr$^{-1}$. 
The formation of the central black hole is however more difficult to explain. Assuming Eddington limited exponential growth, then a 
$10^{9}M_{\odot}$ black hole can grow from a $100M_{\odot}$ `seed' black hole in $\sim725$Myr, however the likelihood that the 
accretion rate is `fine-tuned' to the Eddington limit for many e-foldings appears small, especially considering the role of feedback 
from the formation of stars in the host galaxy \citep{bur01}. It seems therefore that, unless the accretion rate exceeded the Eddington 
limit for some period of time, or the QSO luminosity currently exceeds the Eddington limit, accretion onto an initially stellar mass 
black hole is unlikely to produce the central black hole in SDSS J1030+0524 within the required timescale, and that a significant fraction 
of the black hole mass must be built via some other mechanism. We briefly mention two of a variety of possibilities \citep{ree84,hqu04} here. 
The first is that a more massive seed black hole could form as result of collision runaway in dense young star clusters \citep{por04,gfr04}, 
which can produce `intermediate' mass black holes of several thousand solar masses on rapid timescales. Another possibility is that mergers 
between $\sim100$M$_{\odot}$ black holes created in supernovae of high-mass population III stars contribute to the build-up of more 
massive black holes. Numerical simulations \citep{abn02} suggest that one $\sim100$M$_{\odot}$ star could form rapidly in each 
$\sim10^{6}$M$_{\odot}$ $\Lambda$CDM `minihalo'. Population III stars are not expected to lose much mass in the final stages of stellar 
evolution and may thus be expected to form black holes  of $\sim100$M$_{\odot}$ in less than 1Myr. A fraction of these black holes are 
predicted to be driven to the inner regions of larger galaxies by ongoing mergers \citep{mre01}. The space density and the further fate 
of these black holes is uncertain, but it is at least plausible that mergers between these black holes first form `intermediate' mass 
black holes which then  contribute to the build-up of the $\sim10^{9}$M$_{\odot}$ black holes in $z>6$ QSOs by a mixture of further 
merging and accretion \citep{vol03}. 

In summary, there is observational and theoretical evidence that the host halo, stellar mass, and central black hole required 
to make a fully formed QSO can form in less than a Gyr. The existence however of a radio quiet QSO at $z=6.3$ that appears indistinguishable 
from other RQQs at lower redshifts is still surprising, and the most pressing, and as yet unanswered question is whether the formation 
of the halo, stars, and central black hole in an object such as SDSS J1030+0524 can be accomplished {\it together} in less than a Gyr in the 
$\Lambda$CDM framework.

\acknowledgments
We thank Mark Lacy, Carol Lonsdale, Kirpal Nandra and Mike Norman for helpful discussion, and the referee for a very useful report. 
This paper is based on observations obtained with XMM-Newton, an ESA science mission
with instruments and contributions directly funded by ESA Member
States and the USA (NASA). This research has made use of data  
from the High Energy Astrophysics Science Archive Research Center (HEASARC), 
provided by NASA's Goddard Space Flight Center, and of the NASA/IPAC 
Extragalactic Database (NED) which is operated by the Jet Propulsion 
Laboratory, California Institute of Technology, under contract 
with NASA.


\begin{deluxetable}{lcccccccc}
\tablecolumns{8}
\tablewidth{0pc}
\tablecaption{Results from fitting the X-ray spectrum of SDSS J1030+0524 \label{specfit}}
\tablehead{
\colhead{Model}&\colhead{$\chi^{2}_{\rm red}$/DOF}&\colhead{$\Gamma$}&\colhead{$N_{\rm H}$}&\colhead{$f_{\rm 0.5-12}$}&\colhead{$f_{\rm 0.5-2}$}&\colhead{$f_{\rm 2-10}$}&\colhead{$L_{\rm 0.5-2}$}&\colhead{$L_{\rm 2-10}$} \\
                   &         &               &($10^{22}$ cm$^{-2}$)&\multicolumn{3}{c}{($10^{-15}$ ergs cm$^{-2}$ s$^{-1}$)} & \multicolumn{2}{c}{($10^{45}$ ergs s$^{-1}$)}
}
\startdata
Power law                   &1.50/24 & $2.12\pm0.11$ & --            & 13.70 & 6.28 & 6.73 & 0.17 & 2.87  \\
Power law plus              &        &               &               &       &      &      &      &       \\
absorption                  &1.51/23 & $2.27\pm0.19$ & $2.78\pm2.97$ & 12.98 & 6.49 & 5.96 & 0.04 & 2.75  \\
$\Gamma=2$ power law        &        &               &               &       &      &      &      &       \\
plus absorption             &1.57/24 & {\bf 2.0}     & $0.04\pm1.66$ & 15.0  & 6.19 & 7.92 & 0.14 & 2.68  \\
Power law plus              &        &               &               &       &      &      &      &       \\
fixed absorption            &1.48/24 & $2.18\pm0.11$ & {\bf 1.0}     & 13.38 & 6.36 & 6.40 & 0.10 & 2.82  \\

\enddata
\medskip
 
All fits include a Galactic column of 3.2$\times10^{20}$ cm$^{-2}$. Quantities in bold are fixed during fitting. Fluxes are quoted in the observed 
frame. Luminosities are quoted in the rest frame. 
  
\end{deluxetable}

\begin{thebibliography}{}

\bibitem[Abel, Bryan \& Norman(2002)]{abn02}
Abel T., Bryan G. L., Norman M. L., 2002, Sci, 295, 93

\bibitem[Arnaud(1996)]{arn96}
Arnaud K. A., Astronomical Data Analysis Software and Systems V, A.S.P. Conference Series, 
Vol. 101, 1996, George H. Jacoby and Jeannette Barnes, eds., p. 17.

\bibitem[Barkana \& Loeb(2003)]{blo03}
Barkana R., Loeb A., 2003, Nature, 421, 341

\bibitem[Becker et al(2001)]{bec01}
Becker R. H., et al, 2001, AJ, 122, 2850

\bibitem[Borys et al(2003)]{bor03}
Borys C., Chapman S., Halpern M., Scott D., 2003, MNRAS, 344, 385

\bibitem[Boyle et al(2000)]{boyl00}
Boyle B. J., Shanks T., Croom S. M., Smith R. J., Miller L., 
Loaring N., Heymans C., 2000, MNRAS, 317, 1014

\bibitem[Brandt et al(2002)]{bra01}
Brandt W. N., et al, 2002, ApJ, 569, L5

\bibitem[Burkert \& Silk(2001)]{bur01}
Burkert A., Silk J., 2001, ApJ, 554, L151


\bibitem[Di Matteo et al(2003)]{dim03}
Di Matteo T., Croft R. A. C., Springel V., Hernquist L., 2003, ApJ, 593, 56

\bibitem[Fan et al(2001)]{fan01}
Fan X., et al, 2001, AJ, 122, 2833

\bibitem[Fan et al(2003)]{fan03}
Fan X., et al, 2003, AJ, 125, 1649

\bibitem[Farrah et al(2003)]{far03}
Farrah D., Afonso J., Efstathiou A., Rowan-Robinson M., Fox M., Clements D., 
2003, MNRAS, 343, 585	

\bibitem[Ferrero \& Brinkmann(2003)]{fbr03}
Ferrero E., Brinkmann W., 2003, A\&A, 402, 465

\bibitem[Freudling, Corbin \& Korista(2003)]{fre03}
Freudling W., Corbin M. R., Korista K. T., 2003, ApJ, 587, L67

\bibitem[Fukushige \& Makino(2001)]{fma01}
Fukushige T., Makino J., 2001, ApJ, 557, 533

\bibitem[Gao et al(2004)]{gao04}
Gao L., Loeb A., Peebles P. J. E., White S. D. M., Jenkins A., 
2004, ApJ submitted, astroph 0312499

\bibitem[Grupe et al(2004)]{gru04}
Grupe D., Mathur S., Wilkes B., Elvis M., 2004, AJ, 127, 1

\bibitem[Gurkan, Freitag \& Rasio(2004)]{gfr04}
Gurkan M. A., Freitag M., Rasio F. A., 2004, ApJ, 604, 632

\bibitem[Haehnelt, Natarajan \& Rees(1998)]{hnr98}
Haehnelt M. G., Natarajan P., Rees M. J., 1998, MNRAS, 300, 817

\bibitem[Haiman \& Cen(2002)]{hce02}
Haiman Z., Cen R., 2002, ApJ, 578, 702

\bibitem[Haiman \& Quataert(2004)]{hqu04}
Haiman Z., Quataert E., 2004, astroph 0403225

\bibitem[Isaak et al(2002)]{isa02}
Isaak K. G., Priddey R. S., McMahon R. G., Omont A., Peroux C., Sharp R. G., 
Withington S., 2002, MNRAS, 329, 149

\bibitem[Laor et al(1997)]{lao97}
Laor A., Fiore F., Elvis M., Wilkes B. J., McDowell J. C., 1997, ApJ, 477, 93

\bibitem[Loeb \& Peebles(2003)]{lpe03}
Loeb A., Peebles P. J. E., 2003, ApJ, 589, 29

\bibitem[Madau \& Rees(2001)]{mre01}
Madau P., Rees M. J., 2001, ApJ, 551, L27

\bibitem[Magorrian et al(1998)]{mag98}
Magorrian J., et al, 1998, AJ, 115, 2285

\bibitem[Mathur et al(2002)]{mat02}
Mathur S., Wilkes B. J., Ghosh H., 2002, ApJ, 570, L5

\bibitem[Mo \& White(2002)]{mwh02}
Mo H. J., White S. D. M., 2002, MNRAS, 336, 112

\bibitem[Peacock et al(1998)]{pea98}
Peacock J. A., et al, 1998, MNRAS, 296, 1089

\bibitem[Pentericci et al(2002)]{pen02}
Pentericci L., et al, 2002, AJ, 123, 2151

\bibitem[Petric et al(2003)]{pet03}
Petric A., Carilli C., Bertoldi F., Fan X., Cox P., Strauss M. A., 
Omont A., Schneider D. P., 2003, AJ, 126, 15

\bibitem[Portegies Zwart et al(2004)]{por04}
Portegies Zwart S. F., Baumgardt H., Hut P., Makino J., McMillan S. L. W., 
2004, Nature, 428, 724

\bibitem[Priddey et al(2003)]{pri03}
Priddey R. S., Isaak K. G., McMahon R. G., Robson E. I., Pearson C. P., 
2003, MNRAS, 344, L74

\bibitem[Rees(1984)]{ree84}
Rees M. J., 1984, ARA\&A, 22, 471

\bibitem[Reeves \& Turner(2000)]{rtu00}
Reeves J. N., Turner M. J. L., 2000, MNRAS, 316, 234

\bibitem[Reeves et al(2001)]{ree01}
Reeves J. N., et al, 2001, A\&A, 365, L116

\bibitem[Richards et al(2004)]{ric04}
Richards G. T., et al, 2004, AJ, 127, 1305





\bibitem[Stark et al(1992)]{sta92}
Stark A. A., Gammie C. F., Wilson R. W., Bally J., Linke R. A., 
Heiles C., Hurwitz M., 1992, ApJS, 79, 77



\bibitem[Vignali et al(2003a)]{vig03a}
Vignali C., Brandt W. N., Schneider D. P., Garmire G. P., Kaspi S., 
2003a, AJ, 125, 418

\bibitem[Vignali et al(2003b)]{vig03}
Vignali C., et al, 2003b, AJ, 125, 2876

\bibitem[Volonteri, Haardt \& Madau(2003)]{vol03} 
Volonteri M., Haardt F., Madau P., 2003, ApJ, 582, 559

\bibitem[Walter \& Fink(1993)]{wfi93}
Walter R., Fink H. H., 1993, A\&A, 274, 105
 
\end{thebibliography}
\end{document}